\documentclass[fleqn,twoside]{article}
\usepackage{espcrc2}

\usepackage{graphicx}
\usepackage[figuresright]{rotating}

\newcommand{\AmS}{{\protect\the\textfont2
  A\kern-.1667em\lower.5ex\hbox{M}\kern-.125emS}}

\title{Consequences of
$ t $-channel unitarity for $\gamma^{(*)}p$ and $\gamma^{(*)}\gamma^{(*)}$
amplitudes.}

\author{J.R. Cudell\thanks{J.R.Cudell@ulg.ac.be}, E. Martynov\thanks{E.Martynov@guest.ulg.ac.be}\thanks{on leave from Bogolyubov Institute for Theoretical Physics,
Kiev.} and G. Soyez\thanks{G.Soyez@ulg.ac.be}\\
Universit\'e de Li\`ege, B\^at. B5-a, Sart Tilman, B4000 Li\`ege,
Belgium}
       
\begin{document}

\begin{abstract}
We show that $t$-channel unitarity constraints make it possible to obtain
the photon-photon cross sections from the photon-proton and proton-proton cross
sections, for $Q^2<150$ GeV$^2$. In order to do so, one must postulate the existence of double-pole
or triple-pole singularities in the complex $j$ plane.
\end{abstract}

\maketitle

It is an old result \cite{factorisation} that one can relate the amplitudes
describing three elastic processes $ aa\rightarrow aa $,
$ ab\rightarrow ab $, $ bb\rightarrow bb $. The trick is to continue these 
to the crossed channels  $ a\overline{a}\rightarrow a\overline{a}
$,
$ a\overline{a}\rightarrow b\overline{b} $, $ b\overline{b}
\rightarrow b\overline{b} $, where they exhibit discontinuities 
because of the $a$ and $b$ thresholds. One then obtains a nonlinear system
of equations, which can be solved.  
Working in the complex $j$ plane above thresholds ($t>4m_a^2,\ 4m_b^2$), and defining the matrix 
\begin{equation}
T_{0}=\left( \begin{array}{cc}
A_{aa\rightarrow aa} & A_{ba\rightarrow ba}\\
A_{ab\rightarrow ab} & A_{bb\rightarrow bb}
\end{array}\right) 
\end{equation}
one obtains\begin{equation}
\label{solved}
T_{0}={D\over 1-RD}
\end{equation}
with
$ R_{km}=2i\sqrt{\frac{t-4m_{k}^{2}}{t}} \delta_{km}$
and
$D=T_{0}^{\dagger }$.
The latter is made of the amplitudes on the other side of the cut. For any $D$,
equation (\ref{solved}) is enough to derive factorization: 
the singularities of $T_0$ can only come from the zeroes of 
\begin{equation}
\Delta=\det(1-RD).
\end{equation}
Taking the determinant of both sides of eq. (\ref{solved}), we obtain in the vicinity of $\Delta=0$
\begin{equation}
A_{aa\rightarrow aa}A_{bb\rightarrow bb}-A_{ab\rightarrow ab} A_{ba\rightarrow ba}={C\over\Delta},\end{equation}
where $C$ is regular at the zeroes of $\Delta$.
As the l.h.s. is of order $1/\Delta^2$ we obtain the well-known factorization properties from eqs. (2) and (4):
\\ $\bullet$ The elastic hadronic amplitudes have common singularities;
\\ $\bullet$ At each singularity in the complex
$j$ plane, these amplitudes factorise.

These equations are used to extract relations between the residues of
the singularities, which can be continued back to the direct channel.

We have extended \cite{cms} the above argument including all possible thresholds, both elastic and inelastic. The net effect is to keep the structure (\ref{solved}), but with a matrix $D$ that includes multi-particle thresholds.  
Furthermore, we have shown that one does not need to continue 
the amplitudes from one
side of the cuts to the other, but that the existence of complex 
conjugation for the amplitudes is enough to derive (\ref{solved}). 

Hence there is no doubt that the factorization of amplitudes in the complex
$j$ plane is correct, even when continued to the direct channel. 
For isolated simple poles
one obtains the usual factorization relations for the residues.

If $ A_{pq}(j) $ has coinciding simple and double poles ({\it e.g.} colliding
simples poles at $t=0$), 
\begin{equation}
A_{pq}=\frac{S_{pq}}{j-z}+\frac{D_{pq}}{(j-z)^{2}},\end{equation}
one obtains the new relations
\begin{eqnarray}
D_{11}D_{22}&=&{\left( D_{12}\right) ^{2}},\nonumber\\
\label{double}
D_{11}^2S_{22}&=&{D_{12}(2S_{12}D_{11}-S_{11}D_{12})}.
\end{eqnarray}
In the case of triple poles
\begin{equation}
A_{pq}=\frac{S_{pq}}{j-z}+\frac{D_{pq}}{(j-z)^{2}}
+\frac{F_{pq}}{(j-z)^{3}},\end{equation}
the relations become
\begin{eqnarray}
F_{11}F_{22}&=&{\left( F_{12}\right) ^{2}},\nonumber\\
\label{triple}
F_{11}^2D_{22}&=&{F_{12}(2D_{12}F_{11}-D_{11}F_{12})},\\
F_{11}^3S_{22}&=&F_{11}F_{12}\left( 2S_{12}F_{11}-S_{11}F_{12}\right)
\nonumber\\
&+&D_{12}F_{11}\left( D_{12}F_{11}-2D_{11}F_{12}\right)\nonumber\\
&+&D_{11}^2F_{12}^2.\nonumber
\end{eqnarray}

Although historically one has used $t$-channel unitarity to derive factorization
relations in the case of simple poles, it is now clear \cite{DoLa} that a soft
pomeron pole is not sufficient to reproduce the $\gamma^*$ data from HERA \cite{HERA}. 
However, it is possible, using multiple poles, to account both for the soft
cross sections and for the DIS data \cite{multiple}. We shall see later that
relations (\ref{double}, \ref{triple}) enable us to account
for the DIS photon-photon data from LEP.

For photons, two theoretical possibilities exist:\\
i) The photon cross sections are zero for any fixed number of incoming or
outgoing photons \cite{BN}. In this case, it is impossible to define an
S matrix, and one can only use unitarity relations for the hadronic part
of the photon wave function. Because of this, photon states do not contribute
to the threshold singularities, and the system of equations does not close.
The net effect is that the singularity structure of the photon amplitudes
is less constrained. $\gamma p$ and $\gamma\gamma$ amplitudes must
have the same singularities as the hadronic amplitudes, but extra singularities are possible: in the $\gamma p$ case, 
these may be of perturbative origin, but must have non perturbative residues.
In the $\gamma\gamma$ case, these singularities have their order doubled. It
is also possible for $\gamma\gamma$ to have purely perturbative additional
singularities.

\noindent ii) It may be possible to define collective states in QED for which an S matrix
would exist. In this case, we obtain the same situation
for on-shell photons as for hadrons. However, in the case of DIS, virtual
photons come as external states. Because they are virtual, they
do not contribute to the $t$-channel discontinuities, and hence the singularity
structure for off-shell photons is as described in i).

In the following, we shall explore the possibility that no new singularity
is present for on-shell photon amplitudes, and show that it is in fact
possible to reproduce present data using pomerons with double or triple poles
at $j=1$.

For a given singularity structure, a fit to the $C=+1$ part of proton cross 
sections, and to $\gamma^{(*)}p$ data enables one, via relations (\ref{solved}), 
to predict the $\gamma^{(*)}
\gamma^{(*)}$ cross sections. Hence we have fitted \cite{cms} $pp$ and $\bar
p p$ cross sections and $\rho$ parameters, as well as DIS
data from HERA \cite{HERA}.
\begin{figure}[htb]
\includegraphics*[width=7cm]{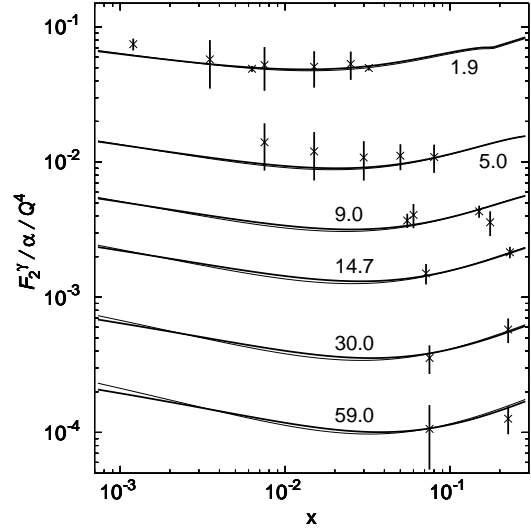}
\caption{Fits to $F_2^\gamma$. The thick and thin curves correspond respectively 
to the triple-pole and to the double-pole cases.  The data are from \cite{LEP}.}
\label{fig:gamma}
\end{figure}

The general form of the parametrisations which we used 
is given, for total cross
sections of $a$ on $b$, by the generic formula
$\sigma^{tot}_{ab}=(R_{ab}+H_{ab})$. 
The first term, from
the highest meson trajectories ($\rho$,
$\omega$, $a$ and $f$),
is parametrized via Regge theory as
\begin{equation} R_{ab}= Y_{ab}^{+} \left({\tilde s}\right)^{\alpha _{+}-1} \pm
Y_{ab}^{-} \left({\tilde s}\right)^{\alpha _{-}-1}\label{lower} \end{equation}
with $\tilde s=2\nu/(1$ GeV$^2)$.
Here the residues $Y_+$ factorise. The second term, from the pomeron,
is parametrized either
as a double pole \cite{DM3}
\begin{eqnarray}
H_{ab}&=& D_{ab}(Q^2)\Re e\left[\log\left(1+\Lambda_{ab}(Q^2)\tilde
s\,^{\delta}\right)\right] \nonumber\\
&+&C_{ab}(Q^2)+(\tilde s\rightarrow -\tilde s)
\label{doubfit}
\end{eqnarray}
or as a triple pole
\begin{equation}  
H_{ab}=t_{ab}(Q^2)\left[ \log^2 \frac{\widetilde{s}}{d_{ab}(Q^2)}
+ c_{ab}(Q^2) \right] .
\label{tripfit}
\end{equation}

The details of the form factors entering (\ref{doubfit}, \ref{tripfit}) can
be found in \cite{cms}.
Such parametrisations give  $\chi^2/dof$ values less than 1.05 
in the region
$\cos(\vartheta_t) \ge  \frac{49}{2m_p^2},\ 
\sqrt{2\nu}\ge 7\ \rm GeV,\ x\leq 0.3,$\break$  Q^2\leq 150 \ {\rm GeV}^2$.

What is really new is that these forms can be extended to photon-photon
scattering, using relations (\ref{double}, \ref{triple}). The total
$\gamma\gamma$ cross section is well reproduced and the de-convolution
using PHOJET is preferred. 
For
photon structure functions, one needs to add one singularity at $j=0$ corresponding
to the box diagram \cite{Budnev}, but otherwise the $\gamma\gamma$ amplitude
is fully specified by the factorization relations. We see in Fig. \ref{fig:gamma} that DIS data are well reproduced by both parametrisations.
\begin{figure}[htb]
\includegraphics*[width=4cm]{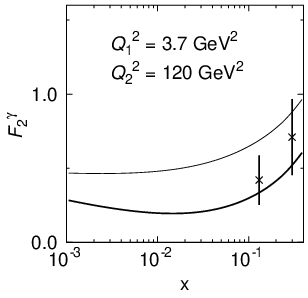}
\includegraphics*[width=7cm]{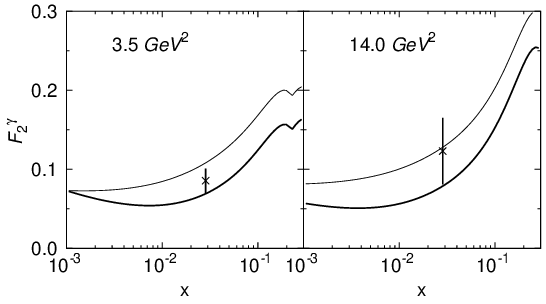}
\caption{Fits to $F_2^\gamma$ for  nonzero asymmetric values of $P^2$ and
$Q^2$ and for $P^2=Q^2$. The curves are as in Fig.~\ref{fig:gamma}.
The data are from \cite{LEP}.}
\label{fig:gsame}
\end{figure}

Even more surprisingly, it is possible to reproduce the $\gamma^*
\gamma^*$ cross sections when both photons are off-shell, as shown in Fig.~\ref{fig:gsame}. This is the place where BFKL singularities may
manifest themselves, but as can be seen such singularities are not needed.

In conclusion, we have shown that it is possible to reproduce soft data ({\it e.g.} total
cross sections) and hard data ({\it e.g.} $F_2$ at large $Q^2$) using a common $j$-plane 
singularity structure, provided the latter is more complicated than simple poles. Furthermore, we have shown that it is then possible to predict $\gamma\gamma$
data using $t$-channel unitarity. How to reconcile such a simple description
with DGLAP evolution, or BFKL results, remains a challenge.

\end{document}